\documentstyle[epsfig]{aipproc}

\def\be{\begin{equation}}
\def\ee{\end{equation}}
\def\ba{\begin{eqnarray}}
\def\ea{\end{eqnarray}}
\def\go{\mathrel{\raise.3ex\hbox{$>$}\mkern-14mu
             \lower0.6ex\hbox{$\sim$}}}
\def\lo{\mathrel{\raise.3ex\hbox{$<$}\mkern-14mu
             \lower0.6ex\hbox{$\sim$}}}
\def\tmev{T_{\rm MeV}}
\def\cm{{\rm cm}}

\begin{document}
\title{Secular Bar-Mode Evolution and
Gravitational Waves From Neutron Stars}

\author{Dong Lai}
\address{Center for Radiophysics and Space Research,
Department of Astronomy\\
Cornell University,
Ithaca, NY 14853}

\maketitle

\begin{abstract}
The secular instability and nonlinear evolution of the $m=2$ f-mode
(bar-mode) driven by gravitational radiation reaction in a rapidly 
rotating, newly formed neutron star are reviewed. There are two types 
of rotating bars which generate quite different gravitational waveforms: 
those with large internal rotation relative to the bar figure 
(Dedekind-like bars) have GW frequency sweeping downward during the evolution,
and those with small internal rotation (Jacobi-like) have GW frequency sweeping
upward. Various sources of viscosity (which affects the instability) 
in hot nuclear matter are reexamined, and the possible effect of
star--disk coupling on the bar-mode instability is also discussed.
\end{abstract}

\section{Introduction}

Nonaxisymmetric instabilities can develop
in rapidly rotating fluid bodies when the ratio $\beta\equiv T/|W|$
of the rotational energy $T$ to the gravitational potential energy $W$
is sufficiently large (e.g., Chandrasekhar 1969).
In particular, the $l=m=2$ f-mode (Kelvin mode), or bar-mode, 
becomes dynamically unstable when $\beta >\beta_{\rm dyn}\simeq 0.27$.  
This $\beta_{\rm dyn}$, originally derived for incompressible 
Maclaurin spheroid, is relatively insensitive to the equation of state 
and differential rotation (Pickett et al.~1996; Toman et al.~1998),
although it tends to be reduced by general relativity (Shibata et al.~2000).
The consequence of the dynamical bar-mode instability has
been extensively studied using numerical simulations: the 
mode grows to nonlinear amplitude by shedding mass and angular momentum
from the ends of the bar in the form of two-armed spiral pattern,
and the central star assumes a bar shape that lasts many rotation periods
(e.g., Tohline et al.~1985; New et al.~2000;
Brown 2000; Shibata et al.~2000). It is important to note, however, that these
simulations start out with a stationary, dynamically unstable star and such an
initial condition may not be realized in an actual core collapse (e.g., 
Rampp, M\"uller and Ruffert 1998). 

However, it is very likely that in a rotating core collapse (or neutron star
binary merger), after the messy dynamics is completed, the newly formed
NS settles down into a dynamically stable configuration, with
$\beta<0.27$, and yet still suffers further secular, rotational
instability. ``Secularly unstable'' means the instability is slow, and is
driven by gravitational radiation reaction. In this paper, we will focus
this secular instability and the possible gravitational wave signals
from newly formed neutron stars in the first seconds/minutes of their lives.

\section{CFS Instability: F-, G- and R-Modes}

The gravitational radiation driven instability, or
Chandrasekhar-Friedman-Schutz (CFS) instability (see e.g., Friedman 1998
for a review), arises for the following simple reason: 
Suppose in the rotating frame of the star we set up a perturbation (mode) 
(with frequency $\omega_r$ as in $e^{im\phi+i\omega_r t}$) which
travels opposite to the rotation. In the inertial frame, the perturbation
will be dragged backward by the rotation, and the mode
frequency becomes $\omega_i=\omega_r-m\Omega_s$. If the spin $\Omega_s$ is 
sufficiently large, the perturbation will be prograde in the inertial frame.
Since the mode has negative angular momentum (because the perturbed fluid
does not rotate as fast as it did without the perturbation), as the
mode radiates positive $J$ through gravitational radiation, the mode's angular
momentum will be more negative; that means the mode is unstable. 

\begin{figure}[b!] 
\vspace{-10pt}
\centerline{\epsfig{file=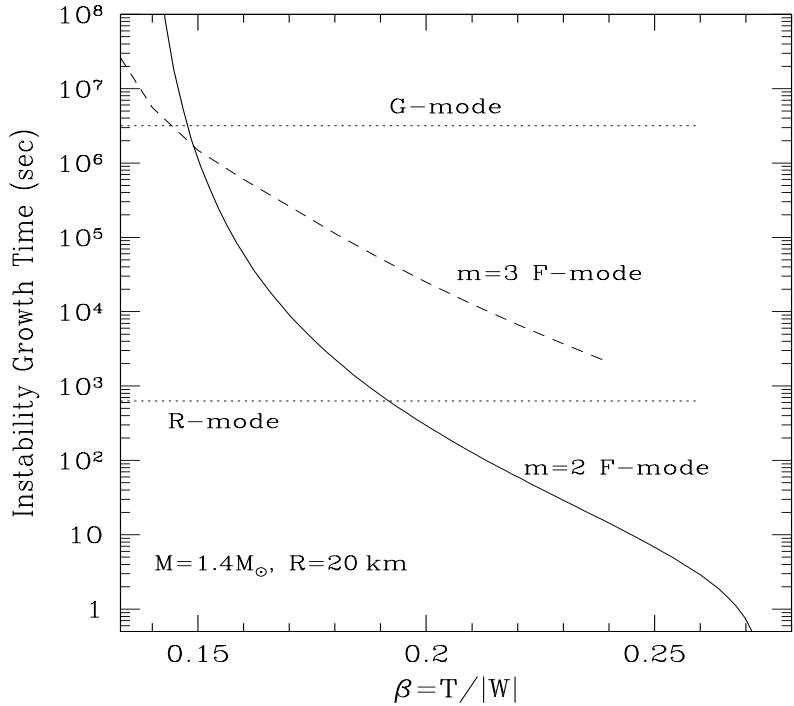,height=4in,width=4.8in}}
\vspace{-0pt}
\caption{Growth times of gravitational radiation driven CFS instability
for different modes in a neutron star as a function of $\beta$.}
\label{fig1}
\end{figure}

The CFS instability mechanism can be applied to different modes of NSs:

{\bf 1. F-modes:} The f-modes are fundamental acoustic waves, corresponding to
a global distortion of the star. For the $m=2$ mode (bar-mode), the
instability occurs when $\beta>\beta_{\rm sec}=0.14$ --- This critical
$\beta_{\rm sec}$ is only slightly affected by equation of state,
and is somewhat reduced (to as small as $0.1$) by strong differential 
rotation (Imamura et al.~1995) and by general relativity 
(Stergioulas \& Friedman 1998).
  
{\bf 2. R-modes:} In the last few years it was realized that the so-called
r-modes are always secularly unstable (for stars consisting of 
inviscid fluid) (Andersson 1998) and the growth time of the mode through
current quadrupole gravitational radiation can be interestingly short
(Lindblom et al.~1998). A lot of works are currently being done on r-modes
(see http://online.itp.ucsb.edu/online/neustars00/si-rmode-sched.html
for a recent meeting devoted to r-modes), but they are beyond the scope of this
paper (see Ushomirsky's paper in this proceedings).

{\bf 3. G-modes:} G-modes are also unstable when $\Omega_s$ is (approximately)
greater than the mode frequency (of nonrotating stars), which typically
is $\sim 100$~Hz, or $0.1\Omega_{\rm max}$. However, the growth time
(despite the finite quadrupole moment) is quite long, an thus not 
interesting (Lai 1999). 

Figure 1 shows a comparison of the CFS instability growth times for different
modes in a $M=1.4M_\odot$, $R_0=20$~km neutron star. 
We see that for sufficiently large $\beta$, the $m=2$ f-mode (the bar-mode) is
the most unstable mode, with the shortest growth time (of order seconds to
minutes). So the bar-mode instability is most robust if a neutron star is
formed with $\beta$ in the range of $0.2$ to $0.27$ --- numerical simulations
indicate that this is indeed possible (Zwerger \& M\"uller 1997; Rampp et
al.~1998; see contributions by New and Brown). In the remainder of the paper we
will be concerned with the bar-mode only.
 
The growth time shown in Fig.~1 is due to gravitational radiation only. There
are several complications that can affect the net growth time. We will
discuss two of them in the next two sections. Other issues,
including the effect of magnetic field, can be found in Ho \& Lai (2000).

\section{Viscosity in Proto-Neutron Stars}

The first complication concerns the viscous dissipation, which tends to
suppress the GR driven instability. For a proto-NS
with $T\go 1$~MeV, the shear (kinematic) viscosity 
due to neutron-neutron scattering, 
$\nu=\eta/\rho
\simeq 14\,\rho_{15}^{5/4}\tmev^{-2}~\cm^2s^{-1}$ (Flowers \& Itoh 1976),
is negligible (since $t_{\rm visc}\sim R^2/\nu$ is much greater than the growth
time of the mode, $t_{\rm GR}$). The shear viscosity due to neutrino-nucleon
scattering is $\eta=n_\nu p_\nu l_\nu/5$ (Goodwin \& Pethick 1982), where 
$n_\nu,p_\nu,l_\nu$ are the neutrino number density, momentum and scattering
mean free path. Thus
\be
\nu={\eta\over \rho}\sim c\, l_\nu\left({E_\nu n_\nu\over \rho c^2}\right)
\sim 3\times 10^6\rho_{15}^{-4/3}\tmev~\cm^2s^{-1},
\ee
where we have used $l_\nu\simeq 2000 (E_\nu/30\,{\rm MeV})^{-3}$~cm.
Contrary to some earlier claims (e.g., Lindblom \& Detweiler 1979),
the neutrino shear viscosity is also negligible. 

The most relevant viscosity in a proto-NS is the bulk viscosity.
This arises from the fact that in an oscillation which involves compression,
the matter will be temporarily out of chemical equilibrium
(in this case, $\beta$-equilibrium between n,p,e) and it will 
try to relax back to the equilibrium by the weak processes 
\be
e+p\rightarrow n+\nu_e,\quad
n\rightarrow p+e+\bar\nu_e,\qquad ({\rm URCA})
\ee
or
\be
e+p+N\rightarrow n+N+\nu_e,\quad
n+N\rightarrow p+N+e+\bar\nu_e,\qquad
({\rm Modified~~ URCA})
\ee
and therefore emitting neutrinos; this neutrino emission serves as damping
of the oscillation.

The standard bulk viscosity widely used in the last decade has been the one
derived by Sawyer (1989) (correcting a factor of 100 typographic error)
\be
\zeta=1.5\times 10^{32}\rho_{15}^2\,\tmev^6\,\omega^{-2}~{\rm g/(cm\,s)},
\qquad
({\rm Modified~~ URCA})
\label{zeta1}\ee
where $\tmev$ is the temperature in MeV, $\rho=10^{15}\rho_{15}$~g/cm$^3$ is
the density, and $\omega$ is the mode (angular) frequency in $s^{-1}$.
This is quite large, with the corresponding damping time
$R^2\rho/\zeta\sim 10\,\tmev^{-6}$~s for $\omega\sim 10^3$~s$^{-1}$.
Thus one may conclude that the bar-mode instability is suppressed for $\tmev
\go 1$ (see Ipser \& Lindblom 1991 for more detailed computations). Similarly,
if direct URCA process operates, equation (\ref{zeta1}) should be
replaced by
\be
\zeta\simeq 1.5\times 10^{36}\rho_{15}^2\,\tmev^4\,\omega^{-2}~{\rm g/(cm\,s)}
\qquad ({\rm URCA})
\label{zeta2}\ee
where we have used the free npe gas relation $n_e/n_0=0.0765\,\rho_{15}^2$
($n_0=0.16$~fm$^{-3}$) to evaluate the equilibrium electron density (this is
not consistent since in a free npe gas URCA precesses are suppressed;
but it is adequate for estimate). 
Equation (\ref{zeta2}) is valid for $\omega\gg 118\,x^{-2/3}\tmev^4$~s$^{-1}$
(where the electron fraction $x=n_e/n=0.02\rho_{15}$).

However, one should be careful when using eqs.~(\ref{zeta1})-(\ref{zeta2}) 
at high temperatures. In fact, when $T$ becomes sufficiently large, the bulk
viscosity must go down. The reason is that as $T$ increases, the timescale to 
relax back to $\beta$-equilibrium becomes shorter than the oscillation
period. So the matter will stay very close to $\beta$-equilibrium
during the oscillation and there is very little extra neutrino 
emission associated with the oscillation.
Another effect that needs to be included at high temperatures
is neutrino absorption (such as $\nu_e+n\rightarrow e+p$) which 
also helps to speed up relaxation to $\beta$-equilibrium.

\begin{figure}[b!] 
\vspace{-10pt}
\centerline{\epsfig{file=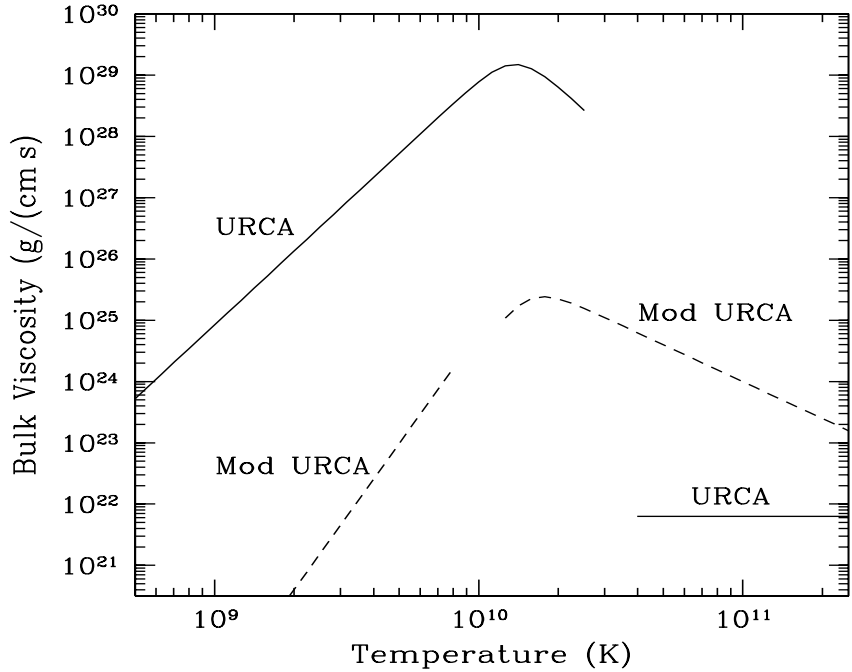,height=4in,width=4.8in}}
\vspace{0pt}
\caption{Bulk viscosity $\zeta$ as a function of temperature.
The lower temperature segments correspond to (neutrino)
optically thin regime, and the high temperature segments to
the opaque regime. The density is $\rho_{15}=1$, and the mode
frequency $\omega/(2\pi)=10^3$~Hz.}
\label{fig2}
\end{figure}

We now outline the derivation of the neutrino bulk viscosity in the regime 
where matter is opaque to neutrinos (see Sawyer 1980; Lai 2001). 
Consider the emission and absorption
of $\nu_e$'s. The $\nu_e$ distribution function $f$ satisfies
the Boltzmann equation
\be
{\partial f\over \partial t}=j-\kappa f,
\ee
where $j$ is the emissivity, and 
$\kappa=j\left(1+e^{E-\delta\mu\over T}\right)$
is the absorption cross section per unit volume (corrected for
the effect of stimulated absorption), $\delta\mu\equiv
\mu_e+\mu_p-\mu_n$.
In equilibrium ($\delta\mu=0$; we assume $\mu_\nu=0$),
we have $f=f_0=f_{eq}=\left(1+e^{E\over T}\right)^{-1}$ and
$j_0=\kappa_0 f_0$. In a perturbation with $\delta f\propto e^{i\omega t}$,
we find
\be
(i\omega +\kappa_0)\delta f={e^{E/T}\over (1+e^{E/T})^2}\left({\delta\mu
\over T}\right)\kappa_0.
\ee
Similar consideration for $\bar\nu_e$ gives
$(i\omega +\kappa_0)\delta\bar f=-{e^{E/T}(1+e^{E/T})^{-2}}\left({\delta\mu
/T}\right)\kappa_0$,
where we have used $\bar j_0=j_0$ and $\bar\kappa_0=\kappa_0$. The variation
of electron fraction $x=n_e/n$ is 
\be
\delta x=-{1\over n}\int\!{d^3p\over (2\pi)^3}\left(\delta f-\delta\bar f
\right)\simeq -{\delta\mu\over (i\omega+\kappa_0)n}
\lambda_{\rm eff},
\ee
where
$\lambda_{\rm eff}=T^2\kappa_0/6$.
Since the matter is degenerate to a good approximation, 
$\delta\mu$ depends only on $\rho,x$. Thus
\be
\delta x=-{\lambda_{\rm eff}/\rho\over i\omega+\kappa_{\rm eff}}
\left({\partial\delta\mu\over\delta n}\right)_x\delta\rho,
\label{delx}\ee
where
$\kappa_{\rm eff}=\kappa_0+({\lambda_{\rm eff}/n})
\left({\partial\delta\mu/\delta x}\right)_n$.
The energy dissipation rate per unit volume (averaged over oscillation period)
can be calculated by 
\be
\langle\dot E\rangle={1\over\rho}\left\langle \delta P {d\delta\rho\over dt}
\right\rangle
={1\over 2\rho}(\delta\rho)^2
{\rm Re}\left[(-i\omega){\delta P\over\delta\rho}\right]
\equiv {1\over 2}\zeta\omega^2\left({\delta\rho\over\rho}\right)^2.
\ee
Using eq.~(\ref{delx}) we then find the bulk viscosity
\be
\zeta={\lambda_{\rm eff}\over \omega^2+\kappa_{\rm eff}^2}
\left({\partial\delta\mu\over\partial n}\right)_x\left({\partial P\over
\partial x}\right)_\rho=
{\lambda_{\rm eff}\over \omega^2+\kappa_{\rm eff}^2}
\left({\mu_n\over 3}\right)^2,
\ee
where the second equality applies for free npe gas ($\mu_n=
140\,\rho_{15}^{2/3}$~MeV is the neutron Fermi energy), for which
$(\partial\delta\mu/\partial x)_n=3(n/n_e)\mu_n$ and thus $\kappa_{\rm eff}
\simeq \kappa_0$.

For modified URCA process, we have
\ba
&&\kappa_0=1.034\times 10^3\,\rho_{15}^{2/3}\tmev^4~{\rm s}^{-1},\\
&&\zeta=7.9\times 10^{31}\,\rho_{15}^2\,\tmev^6\,(\omega^2+\kappa_0^2)^{-1}
~{\rm g/(cm\,s)}.\qquad
({\rm Modified~~ URCA})
\ea
If direct URCA process operates (which requires sufficiently large $x$, 
a condition necessarily satisfied at early times of the 
proto-neutron star), we have
\ba
&&\kappa_0=1.13\times 10^7\,\rho_{15}^{2/3}\tmev^2~{\rm s}^{-1},\\
&&\zeta=6.7\times 10^{21}\,\rho_{15}^{2/3}~{\rm g/(cm\,s)},\qquad
({\rm URCA})
\ea
where the $\zeta$ expression is evaluated in the $\kappa_0\gg\omega$ limit.
We see from Fig.~2 that for $T\go$~MeV, the viscosity is less than 
$10^{25}$~g/(cm~s) (depending on whether URCA or modified URCA
processes operates), and thus the bulk viscosity cannot suppress the CFS
instability of the bar-mode (see Lai 2001 for more details).

\section{Effect of Star--Disk Coupling}

A newly formed rotating neutron star is often surrounded by a disk.
As the bar grows in the NS, it will excite density waves in the disk,
and therefore transfers angular momentum to the disk or from
the disk. This will either enhance or suppress the bar-mode
instability --- this is simply another form of CFS instability.

The angular momentum transfer is mainly through the so-called Lindblad 
resonances, where $2(\Omega_p-\Omega_k)=\pm\Omega_k$ ($\Omega_p=\omega_i/2$
is the pattern rotation of the bar and $\Omega_k$ is the
rotation of the disk, assumed to be Keplerian). 
The torques can be calculated using the formalism of Goldreich \& Tremaine
(1979). At the inner Lindblad resonance (ILR), $\Omega_k=2\Omega_p$, 
the driving rate of the bar-mode due to star--disk coupling is (Lai 2001)
\be
\gamma_{\rm ILR}\simeq -{6\pi^2\over 5}{\Omega_p^2\over(\Omega_s-2\Omega_p)}
{\Sigma(r_{\rm ILR})\over (M/R_e^2)},
\ee
where the negative sign in the front implies that the torque tends to damp the
mode, $\Omega_s$ is the rotation rate of the star, $R_e$ is the equatorial
radius, and $\Sigma(r_{\rm ILR})$ is the surface density of the disk 
evaluated at ILR. Similarly, at the outer Lindblad resonance (OLR), we have
\be
\gamma_{\rm OLR}\simeq {98\pi^2\over 45}{\Omega_p^2\over(\Omega_s-2\Omega_p)}
{\Sigma(r_{\rm OLR})\over (M/R_e^2)},
\ee
where the positive front sign means that the torque drives the CFS instability.
Clearly, the net effect of star-disk coupling on the mode depends on 
the relative importance of ILR and OLR. Compared with driving rate of the mode
due to gravitational radiation, we find that star-disk coupling
is important when $\Sigma\go 10^{-6}M/R_e^2$. So even a small
amount of material outside the proto-NS may potentially affect the 
CFS instability of the bar-mode.

\section{Nonlinear Evolution of Bar-Mode and Gravitational Waveforms}

\begin{figure}[b!] 
\vspace{-30pt}
\centerline{\epsfig{file=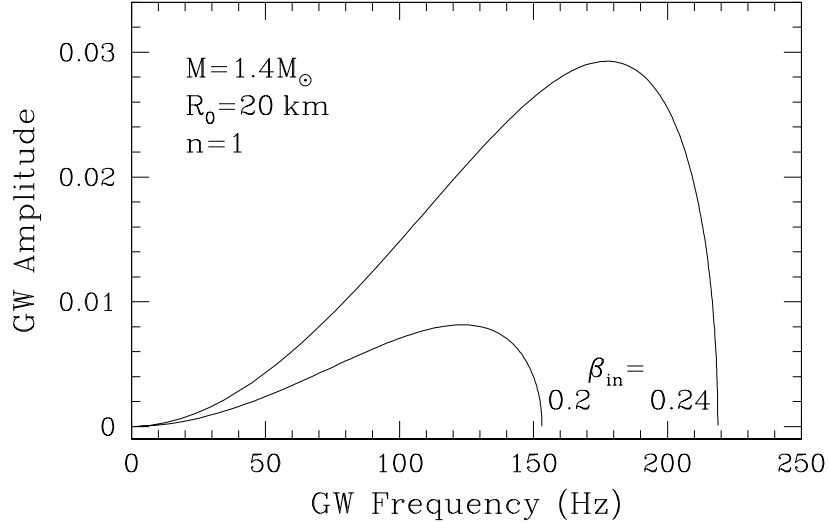,height=5in,width=5in}}
\vspace{-120pt}
\caption{The amplitudes of the GWs emitted by a secularly unstable NS, 
evolving from a Maclaurin spheroid toward a Dedekind ellipsoid. 
The two curves correspond to initial $\beta=0.24$ and $0.2$ respectively
($n=1$ is the polytropic index).
The GW frequency sweeps downward.}
\label{fig3}
\end{figure}

In general, to determine the nonlinear evolution of the bar-mode
requires 3d hydrodynamical simulations including 
gravitational radiation reaction, and one needs to follow the system 
for a time much longer than the dynamical time of the star.
This possesses a significant technical challenge (see recent attempt
by Lindblom et al.~2000 on the r-mode evolution where an approximate
ansatz for the radiation reaction is adopted). 

\begin{figure}[b!] 
\vspace{-30pt}
\centerline{\epsfig{file=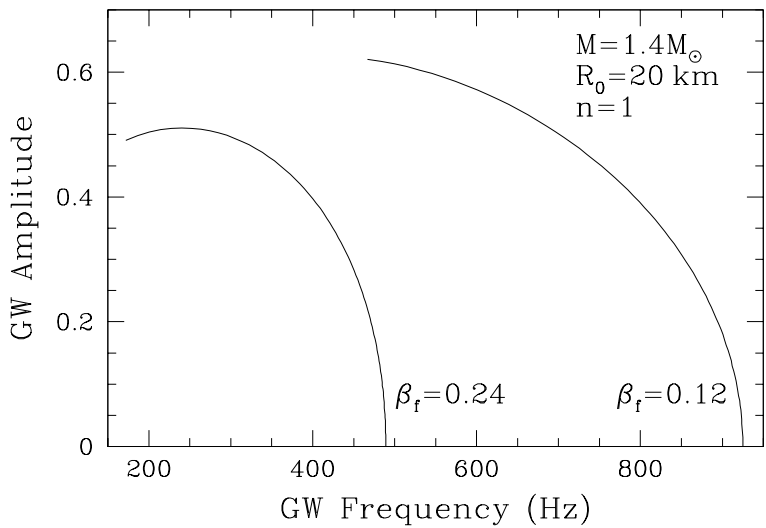,height=5in,width=5in}}
\vspace{-120pt}
\caption{The amplitudes of the GWs emitted by Jacobi-like bars.
The two curves correspond to final $\beta=0.24$ and $0.12$ respectively.
The GW frequency sweeps upward.}
\label{fig4}
\end{figure}

\subsection{Dedekind-like Bar}

One of the advantages of the $m=2$ f-mode (as opposed the higher $m$ f-modes
and or r-modes) is that under certain idealized condition, namely for
impressible fluid, there exists an exact solution for the nonlinear 
development of the bar-mode instability driven by gravitational radiation
(Miller 1974; see Lai \& Shapiro 1995 and references therein; the latter also
includes an approximate, compressible generalization).


The evolutionary sequence is as follows:
We start with an axisymmetric Maclaurin spheroid with $\beta>0.14$ 
(secularly unstable). The bar grows and has a pattern angular 
frequency $\Omega_p$ (which is related to the mode frequency
in the inertial frame by $\Omega_p=\omega_i/2$).
Relative to the bar, there is also an internal
rotation $\Omega_{in}$ which is larger than $\Omega_p$. The important
point to note is that although the mean rotation of the star 
$\Omega_s\simeq \Omega_p+\Omega_{in}$ is near breakup,  
$\Omega_p$ can be much smaller (in fact, at the bifurcation
point $\beta=0.14$, we have $\Omega_p=0$).  As the amplitude
of the bar continues to grow, the bar also gradually slows down.
Eventually we reach a configuration with zero pattern speed.
This is the Dedekind ellipsoid, basically stationary ``football''
with a fixed figure in space but with a lot internal rotation $\Omega_{in}$.

The gravitational wave (GW) emitted during such a quasi-equilibrium
secular evolution is quite interesting. Figure 3 shows the GW 
amplitude as a function of GW frequency. The GW is quasi-periodic.
Initially, we have an axisymmetric star, so $h=0$. Then the GW amplitude
increases as the bar grows. In the meantime, the bar slows down
and the GW frequency ($=2\Omega_p$) decreases. So eventually  
$h$ decreases. Thus we have a non-monotonic GW amplitude evolution,
with the GW frequency sweeps downward from a few hundred Hertz
toward zero. The timescale of the evolution is of order seconds to minutes, 
and the characteristic number of cycles of GWs is of order $10^4$.

So far we have assumed the star begins its evolution from an
axisymmetric state. Recent numerical simulations (e.g., New et al.~2000,;
Brown 2000; Shibata et al.~2000), however, indicate that at the end of
the dynamical evolution, the star may be already elongated rather than
axisymmetric. So one can ask about the long-term evolution of the bar and
the emitted GWs. Here the ellipsoid model can also provide qualitative
answers. There are two possibilities (see Lai \& Shapiro 1995 for 
details): (1) If the bar has $\Omega_{in}>\Omega_p$, namely if the bar is
{\it Dedekind-like}, then the evolution and waveform discussed above
should also apply except that we need to cut off the initial growth
phase of the bar. (2) Another possibility occurs when $\Omega_p>\Omega_{in}$,
which we discuss in the next subsection.

\subsection{Jacobi-like Bar}

If the bar has internal rotation (relative to the bar figure) less than
the pattern rotation, i.e., $\Omega_{in}<\Omega_p$, the bar is called
{\it Jacobi-like}, and the evolution is quite different. In this case, 
gravitational radiation reaction tends to decrease the amplitude of the bar,
making the star less elongated. In the meantime, $\Omega_p$ increases
because, even though $J$ decreases, the moment of inertia decreases faster.
So the GW frequency increases and eventually the star becomes
axisymmetric. Here again we have a quasi-periodic GW signal (see Fig.~4),
except that the frequency sweeps upwards. The timescale foe the evolution
is of order a second, and the number of cycles is of order a few hundred.

Note that at the end of the Jacobi-like evolution, the axisymmetric star may
still be secularly unstable. If so the star will continue to evolve in
the way described in Sec.~V.A.

\subsection{Characteristic GW Amplitude}

\begin{figure}[b!] 
\vspace{-30pt}
\centerline{\epsfig{file=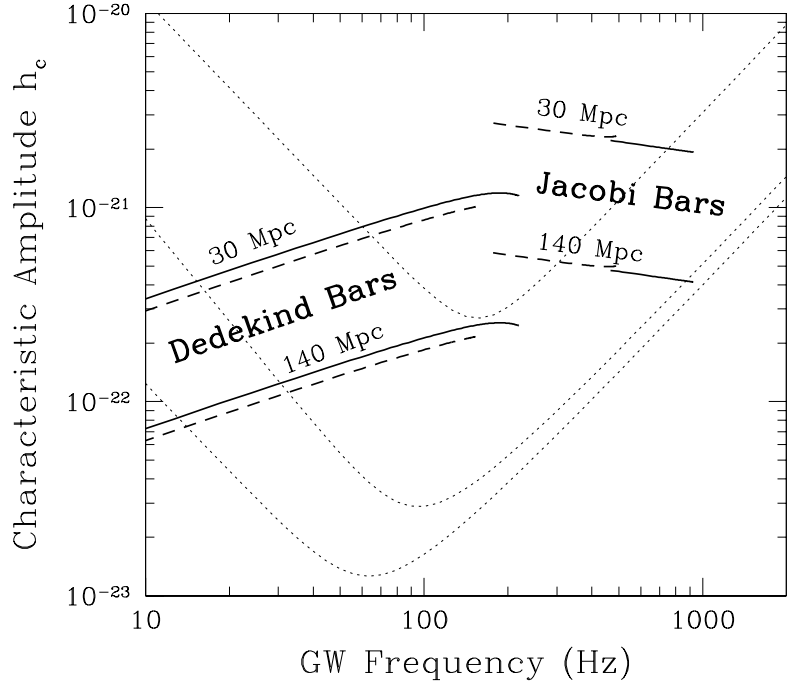,height=5in,width=5in}}
\vspace{-60pt}
\caption{Comparison between the characteristic GW amplitudes $h_c$ emitted
during the secular evolution of a nonaxisymmetric neutron star and the rms
noise $h_{\rm rms}$ of LIGO I,II and III. The solid and dashed lines 
correspond to the two different cases in Fig.~3 and Fig.~4.
}
\label{fig5}
\end{figure}

To recapitulate, there are two types of rotating bars, and their evolution and
emitted GWs are qualitatively different: (1) A Dedekind-like 
bar has large internal rotation, and the resulting waveform sweeps downward in
frequency; (2) A Jacobi-like bar has relatively small internal rotation, and 
the GW frequency sweeps upwards. 
Figure 5 shows the characteristic GW amplitudes for the two types of
bars compared with the sensitivity curves ($h_{\rm rms}$)
of LIGO I,II and III. The characteristic GW amplitude is given by
(Lai \& Shapiro 1995)
\be
h_c=h\left|{dN\over d\ln f}\right|^{1/2}
={M\over D}\left({R_0\over M}\right)^{1/4}\left({5\over 2\pi}
\left|{d\bar E\over d\bar\Omega_p}\right|\right)^{1/2},
\ee
where $D$ is the distance, $|dN/d\ln f|=|f^2/\dot f|$ is the number of cycles
of GW spent near frequency $f$, $E={\bar E}(M^2/R_0)$ is the energy of
the star, and $\Omega_p={\bar\Omega_p}(M/R_0^3)^{1/2}$ is its angular pattern
speed. For a broad band detector such as LIGO, the best
signal-to-noise ratio will be obtained by matched filtering of the data, with 
$S/N\simeq h_c/h_{\rm rms}$. Note that unlike coalescing compact binaries,
the phase evolution of the GW from the evolving bars can not be determined
with the accuracy needed for matched filtering, so such a high $S/N$ may not
be achieved in practice. A new fast chirp transform technique (Jenet \& Prince
2000) is promising for detecting such signals. 

The event rate of Type II SNe at distance of $30$~Mpc is about 100 per year.
So even if a small fraction of the NSs are formed rotating rapidly, 
the GW signals discussed here are promising for lIGO. Of course, 
it should be emphasized that the waveforms discussed in this section
are based on the exact solution for the idealized situation (incompressible
fluid). Whether this idealized solution has any resemblance to reality remains
to be seen by future studies.

\section{Parametrized Waveforms}

Finally we give fitting formulae for the gravitational waveforms
generated by a rapidly rotating NS as it evolves from 
an initial axisymmetric configuration toward a triaxial
ellipsoid (Maclaurin spheroid $\Rightarrow$ Dedekind ellipsoid) 
as discussed in Sec.~V.A. We use units such that $G=c=1$.

The waveform (including the polarization) is given by eq.~(3.6)
of Lai \& Shapiro (1995) (hereafter LS). Since the waveform is quasi-periodic,
we will give fitting formulae for the wave amplitude $h$ (Eq.~[3.7] of LS) 
and the quantity $(dN/d\ln f)$ (eq.~[3.8] of LS; related to the
frequency sweeping rate), from which the waveform $h_+(t)$ and $h_\times(t)$
can be easily generated in a straightforward manner. 

{\bf 1. Wave Amplitude:} The waveform is parametrized by three
numbers: $f_{max}$ is the maximum wave frequency in Hertz,
$M_{1.4}=M/(1.4M_\odot)$ is the NS mass in units of $1.4M_\odot$,
$R_{10}=R/(10\,{\rm km})$ is the NS radius in units of $10$ km. 
(Of course, the distance $D$ enters the expression
trivially.) It is convenient to express the dependence of $h$
on $t$ through $f$ (the wave frequency), with $f(t)$ to be determined 
later. A good fitting formula is 
\be
h[f(t);f_{max},M,R]={M^2\over DR}A\left({f\over
f_{max}}\right)^{2.1}\left(1-{f\over f_{max}}\right)^{0.5},
\label{eq1a}\ee
where 
\be
A=\cases{(\bar f_{max}/1756)^{2.7}, & for $\bar f_{max}\le
400$~Hz;\cr
(\bar f_{max}/1525)^3, & for $\bar f_{max}\ge 400$~Hz;\cr}
~~~~~~{\rm with}~\bar f_{max}\equiv f_{max}R_{10}^{3/2}M_{1.4}^{-1/2}.
\label{eq1b}\ee
Note that if we want real numbers, we have
\be
{M^2\over DR}=4.619\times 10^{-22}M_{1.4}^2R_{10}^{-1}
\left({30\,{\rm Mpc}\over D}\right).
\ee

{\bf 2. Number of GW cycles:} The fitting formula is 
\be
\left |{dN\over d\ln f}\right|=\left({R\over M}\right)^{5/2}
{0.016^2 (R_{10}^{3/2}M_{1.4}^{-1/2} f/1\,{\rm Hz})\over
A^2(f/f_{max})^{4.2}[1-(f/f_{max})]}.
\label{eq2}\ee

\smallskip
\noindent
{\bf Notes to the fitting formulae:}

{\bf Note (i)}: Using the above equations, we obtain the characteristic 
amplitude:
\ba
&&h_c=h\left|{dN\over d\ln f}\right|^{1/2}
=0.016{M^{3/4}R^{1/4}\over D}
\left({R_{10}^{3/2}M_{1.4}^{-1/2}\,f\over 1\,{\rm Hz}}\right)^{1/2}\nonumber\\
&&=5.3\times 10^{-23}\left({30\,{\rm Mpc}\over D}\right)
M_{1.4}^{3/4}R_{10}^{1/4}\left({R_{10}^{3/2}M_{1.4}^{-1/2}
f\over 1\,{\rm Hz}}\right)^{1/2},
\ea
which agrees with Eq.~(3.12) of LS to within $10\%$ [Note that 
in Eq.~(3.12) of LS, the factor $f^{1/2}$ should be replaced by 
$(R_{10}^{3/2}M_{1.4}^{-1/2}\,f)^{1/2}$, similar to the above 
expression.]

{\bf Note (ii)}: The accuracy of these fitting formulae (as compared to 
the numerical results shown in LS) is typically within $10\%$. 
When $f$ is very close to $f_{max}$, the error in the fitting can be
as large as $30\%$.

{\bf Note (iii)}: The frequency evolution $f(t)$ is obtained 
by integrating the equation
$f^2/\dot f=-|dN/d\ln f|$ (note that the frequency sweeps from
$f_{max}$ to zero). For example, we can choose $t=0$ at
$f=0.9f_{max}$. (Note that one should not choose $t=0$ at
$f=f_{max}$ as the time would diverge --- the actual evolution 
near $f_{max}$ depends on the initial perturbations). 
Once $f(t)$ is obtained, the waveform can be calculated as
(cf.~Eq.~[3.6] of LS):
\ba
&&h_+ =h[f(t);f_{max},M,R]\cos\Phi(t){1+\cos^2\theta\over 2},\\
&&h_\times =h[f(t);f_{max},M,R]\sin\Phi(t)\cos\theta,
\ea
where $\theta$ is the angle between the rotation axis of the star 
and the line of sight from the earth, and 
$\Phi(t)=2\pi\int f(t)dt$ is the phase of the gravitational wave. 

{\bf Note (iv)}: $f_{max}$ typically ranges from $100$ Hz to $1000$ Hz
(see Fig.~5 of LS); $M_{1.4}$ and $R_{10}$ are of order unity for 
realistic neutron stars.

\end{document}